\documentclass[a4,12pt,longbibliography,floatfix]{revtex4-2}

\usepackage[utf8]{inputenc} \usepackage[T1]{fontenc}
\usepackage{times}
\usepackage{amsmath}
\usepackage{bm}
\usepackage{amssymb,latexsym,mathrsfs}
\usepackage{url}
\usepackage{color}
\usepackage{tikz}         
\usepackage{xcolor}       
\usepackage{booktabs}
\usepackage{xurl}         
\long\def\comment#1{} 

\usepackage{etoolbox}
\makeatletter
\patchcmd{\frontmatter@RRAP@format}{(}{}{}{}
\patchcmd{\frontmatter@RRAP@format}{)}{}{}{}
\renewcommand\Dated@name{} 
\makeatother

\usepackage[sf,pagestyles]{titlesec}
\titleformat{\section} {\normalfont\sffamily\bfseries} {\thesection}{1em}{}
\titleformat{\subsection} {\normalfont\small\sffamily} {\thesection}{2em}{}

\renewcommand*{\thesection}{\arabic{section}}

\usepackage[singlelinecheck=false,skip=-2mm,labelfont={bf,sf},font=small,justification=raggedright]{caption}

\makeatletter
\def\incepsfig{\@ifnextchar[{\@incepsf}{\@incepsf[tp]}}
\def\@incepsf[#1]#2#3#4{\@ifnextchar[{\@incepsfp[#1]{#2}{#3}{#4}}{\@incepsfp[#1]{#2}{#3}{#4}[]}}
\def\@incepsfp[#1]#2#3#4[#5]{%
\begin{figure*}[#1]%
  \captionsetup{width=160mm}
  \begin{center}%
	  \noindent\hbox to 140mm{\hss\scalebox{#3}{\hss\begin{psfrags}#5{\includegraphics{#2.eps}}\end{psfrags}\hss}\hss}%
  \end{center}%
  \caption{#4}\label{#2}
\end{figure*}}
\makeatother

\usepackage{setspace}
\def\figureref#1{{{Figure~\ref{#1}}}} 
\def\tableref#1{{{Table~\ref{#1}}}} 

\def\cstabhlinedown{\rule[-1.5ex]{0mm}{2ex}} 
\def\cstabhlineup{\rule{0mm}{3.5ex}}

\usepackage{psfrag} 

\let\oldsqrt\sqrt
\def\smallsqrt{\mathpalette\DHLhksqrt}
\def\DHLhksqrt#1#2{\setbox0=\hbox{$#1\oldsqrt{#2\,}$}\dimen0=\ht0
\setbox3=\hbox{\smash{\hbox{$#1\oldsqrt{#2\,}$}}}%
\advance\dimen0-0.2\ht0
\setbox2=\hbox{\vrule width 0.4pt height\ht0 depth -\dimen0}
\setbox4=\hbox{\smash{\hbox{\vrule width 0.4pt height\ht0 depth -\dimen0}}}%
{\box3\lower0.45pt\box4}}

\usepackage[bookmarks=false,breaklinks=true]{hyperref}

\definecolor{royalbluecs}{rgb}{0.2, 0.2, 1}

\hypersetup{linktoc=all, 
    colorlinks=true, linkcolor={royalbluecs},
    citecolor={royalbluecs}, urlcolor={royalbluecs}}

\definecolor{lime}{HTML}{A6CE39}
\DeclareRobustCommand{\orcidicon}{%
	\begin{tikzpicture}
	\draw[lime, fill=lime] (0,0) 
	circle [radius=0.16] 
	node[white] {{\fontfamily{qag}\selectfont \tiny ID}};
	\draw[white, fill=white] (-0.0625,0.095) 
	circle [radius=0.007];
	\end{tikzpicture}
	\hspace{-2mm}}

\foreach \x in {A, ..., Z}{%
	\expandafter\xdef\csname orcid\x\endcsname{\noexpand\href{https://orcid.org/\csname orcidauthor\x\endcsname}{\noexpand\orcidicon}}}

\normalsize\normalfont
\begin{document}

\title{\textsf{On the relation between the three Reidemeister moves \\ and the three gauge groups}}

\author{\ \\ \ \\ 
Christoph Schiller\footnote{{Motion Mountain Research, 
81827 Munich, Germany, \email{cs@motionmountain.net}},
ORCID 0000-0002-8188-6282.} \orcidB{} \\
\ \\ \ \\ }

\begin{abstract} 
\noindent \ \hfil \par 
\noindent {\bf\textsf{Abstract}}\par\noindent
Quantum theory suggests that the three observed gauge groups U(1), SU(2) and SU(3) are related to the three Reidemeister moves: twists, pokes and slides. The background for the relation is provided. It is then shown that twists generate the group U(1), whereas pokes generate SU(2). Emphasis is placed on proving the relation between slides, the Gell-Mann matrices, and the Lie group SU(3). Consequences for unification are deduced.

\bigskip  

\bigskip

\bigskip

\bigskip

\noindent Keywords: Reidemeister moves; gauge groups.

\end{abstract} 

\bigskip

\bigskip

\date{2023}  

\bigskip

\bigskip

\maketitle

\newpage\normalfont\normalsize


\incepsfig{ill-Reide}{0.95}{Three \textit{moves} -- actually shape deformations -- 
for links, knots, tangles and braids were defined by Reidemeister 
\protect\cite{reidemeisternew}: twists, pokes and slides. In the strand tangle model of quantum theory, they 
maintain the topology and thus {maintain the particle type}, but they \textit{change the phase} of the quantum state. Therefore they model interactions.}

\section{The origin of the gauge groups}   
\label{sec:intro}

\noindent
Clarifying the origin of the three gauge groups for the three gauge interactions is an open issue of physics.
Often, the three gauge groups U(1), SU(2) and SU(3) are seen as related to the complex numbers, the quaternions 
and the octonions \cite{baez2002octonions,furey2016standard,singh2020octonions}. 
However, this approach, while successful for U(1) and SU(2), yields SU(3) only indirectly, and only together with additional structures. 
In the following, it is argued that a simpler way to understand the origin of the gauge groups is by relating them 
to the Reidemeister moves. 

The three Reidemeister moves, illustrated in \figureref{ill-Reide}, have been used to 
classify the possible deformations of mathematical links, knots, tangles and braids for almost a century.
The relation between Reidemeister moves and gauge groups that is used in the following arose from Dirac's 
description of spin $1/2$ particles as tethered structures. 
In his lectures, Dirac popularized a variation of the belt trick -- without ever publishing anything 
about it -- that shows how tethered structures behave like spin $1/2$ particles: 
they come back to their original state only after a rotation by $4 \pi$, 
i.e., after \textit{two} full turns  \cite{gardner}. Tethered structures also behave as fermions under exchange.
Continuing this approach, Battey-Pratt and Racey were led to explore tethered particles and the behaviour of their tethers
more thoroughly, and to associate tethers to wave functions.
This association allowed them to deduce the Dirac equation \cite{bpr}.
Extending this approach led to the so-called \textit{strand tangle model}, 
which describes elementary particles as rational tangles of strands and gauge interactions as tangle deformations \cite{cspepan,csorigin,csqed,csqcd}.
The present text focuses on how tangle deformations are related to the generators of nature's gauge groups. 
It will be shown that each of the three Reidemeister moves is related to one of the three 
observed gauge groups which describe the gauge interactions between fermions.


\incepsfig{ill-fundamental-principle}{1}{%
In the strand tangle model, the only observable process is the \textit{local change of sign} of a strand crossing, the so-called \textit{crossing switch}.
This fundamental process generates a quantum of action $\hbar$.
The strands themselves are of Planck radius and thus are \textit{not} observable.
A crossing is a region where the strand distance is minimal.
Only a crossing switch is observable.
As shown in previous papers \protect\cite{csorigin,cspepan,csqed,csqcd,csindian,csbh}, this fundamental principle yields the Lagrangian of the standard model of particle physics, extended with massive Dirac neutrinos and PMNS mixing, and the Lagrangian of general relativity.}[%
\psfrag{t1}{\small $t$}%
\psfrag{t2}{\small $t+\Delta t$}%
\psfrag{WW}{\small $W=\hbar$}%
\psfrag{DL}{\small $\Delta l \geq \sqrt{4\hbar G/c^3}$}%
\psfrag{DK}{\small $\Delta t \geq \sqrt{4\hbar G/c^5}$}%
\psfrag{KK}{\small $S=k\,\ln 2$}%
]
\incepsfig{ill-fermion}{0.95}{%
In the strand tangle model, fluctuating strands of a particle tangle yield wave functions and probability densities. Tangles are fluctuating skeletons of wave functions.}
\incepsfig[ht]{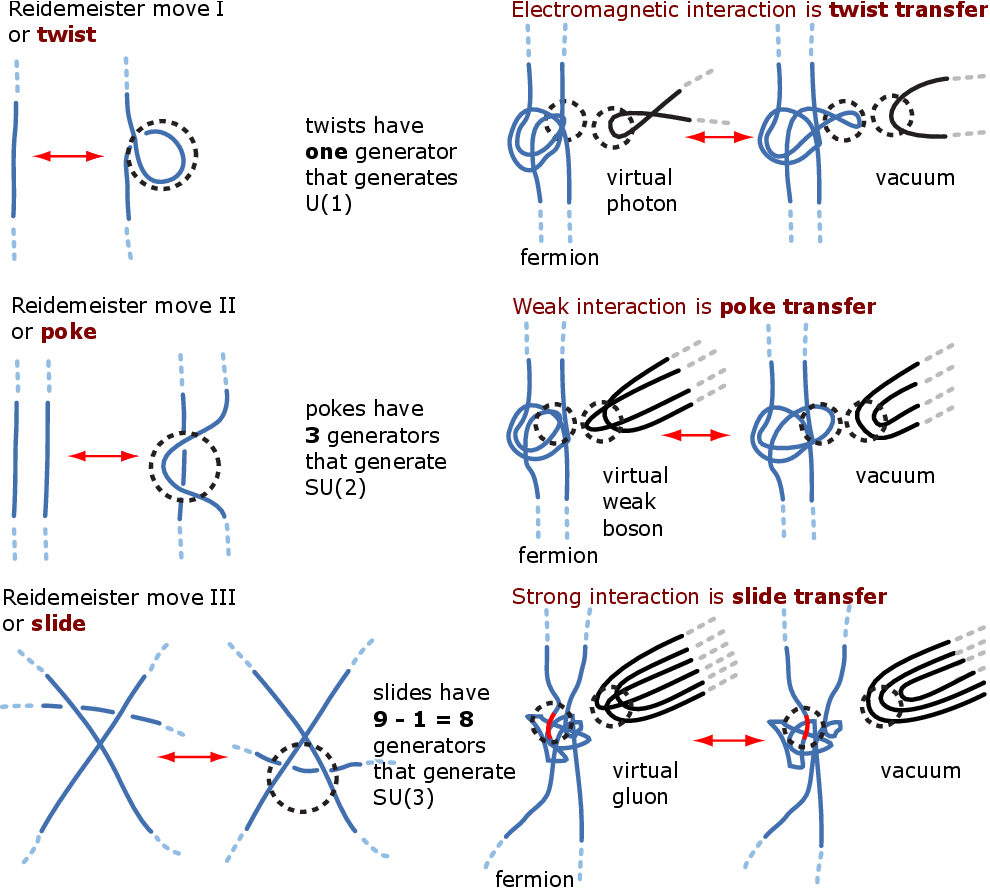}{0.95}{%
The strand tangle model of gauge interactions is based on the theorem that the three Reidemeister moves classify all possible deformations of tangle cores 
\protect\cite{reidemeisternew}.
In particular, the Reidemeister moves determine the generators and the 
elements of the three observed gauge groups, as shown below.
A first, simplified way to describe the Reidemeister moves is by stating that 
they rotate the segment enclosed by a dotted circle by the 
angle $\pi$ -- and then shift it, as shown later on. 
This model yields quantum field theory, as shown in references \protect\cite{cspepan,csorigin,csqed,csqcd}.}

\section{A summary of the strand tangle model} 

\noindent
The present section presents the physical background. 
%
In nature, gauge interactions change the state of fermions: \textit{gauge interactions change the phase of particle wave functions.}
To explore the relation between gauge groups and Reidemeister moves,
it is thus necessary to clarify how strands lead to particles, to their phase, and to wave functions.
This was done extensively by Battey-Pratt and Racey \cite{bpr} and in previous articles \cite{cspepan,csorigin,csqed,csqcd}, all based on 
Dirac's explanation of spin $1/2$ particles using tethered structures.

In the strand tangle model, particles are rational tangles of \textit{strands,} as illustrated in \figureref{ill-fermion}.
{Strands} are fluctuating lines -- actually of Planck radius -- in space, with no ends, that reach the cosmological horizon.
Because of the minimum length in nature, strands are impenetrable, uncuttable and unobservable.
The only observable process occurs when, in a region in space, a strand segment that was \textit{behind} another strand, 
suddenly is \textit{in front} of the same strand. 
This local process 
-- which slightly depends on the observer -- 
is called a \textit{crossing switch}.
In the strand tangle model, a crossing switch generates a quantum of action $\hbar$.
This statement, illustrated in \figureref{ill-fundamental-principle}, is the fundamental principle of the strand tangle model. 
Every crossing switch is observable, and every observable is due to crossing switches.

All particles are rational \textit{tangles} of strands.
Therefore, all particles are tethered structures.
The \textit{tethers} are those strand segments that reach the
cosmological horizon. 
The tangle topology determines the \textit{type} of the particle -- i.e., the \textit{substance}. 
For example, the conservation of topology over time reproduces the conservation of particle type.
The average geometrical \textit{shape} -- i.e., the average {form} -- determines the \textit{state} of the particle.
In particular, the (average) location of the tangled part, the \textit{tangle core}, determines the \textit{position} of the particle. 
The \textit{phase} of a wave function is the (average) orientation of the tangle core.
The \textit{wave function} arises through the \textit{blurring} (i.e., the averaging) of the fluctuating tangle crossings. 
Conversely, the tangle is the fluctuating \textit{skeleton} of a wave function.
The derivation of the probability density from a tangle is briefly illustrated in \figureref{ill-fermion}. 
Spinors and antiparticles can also be deduced.

Particles are \textit{rational} tangles of strands.
Rational tangles are \textit{unknotted} tangles that can be untied by moving the tethers around.
Braids are typical examples of rational tangles.
Elementary particles are rational tangles consisting of one, two or three strands.
Classifying such rational tangles leads to the observed spectrum and quantum numbers of elementary particles \cite{cspepan,csorigin,csqed,csqcd}.

In the strand tangle model, gauge interactions are tangle \textit{core deformations}.
Core deformations lead to a phase change for the particle.
About a century ago, Reidemeister proved that all core deformations that maintain tangle topology can be 
composed of just three basic types, illustrated in \figureref{ill-Reide}, which he called \textit{moves} \cite{reidemeisternew}. 
The general relation between the three moves and the gauge interactions is summarized in \figureref{ill-gauge}.
The core deformation process leads to a model for the emission and absorption of gauge bosons; 
in turn, this process allows reproducing perturbative quantum field theory \cite{csqed,csqcd}.

In short, tangles of unobservable strands with observable crossing switches reproduce spin $1/2$, wave functions, phase, probability densities, spinors, and quantum (field) theory.
As will be shown now, the strand tangle model reproduces the three gauge groups.


\incepsfig{ill-twist-u1}{1}{%
Top: Twists, the first Reidemeister moves, can be described as the local rotation by $\pi$ of an encircled (red) strand segment around a given (dotted) axis.
The dotted circle can be imagined to be the border of a transparent plastic disc glued to the segment, or a fist from the left grabbing the segment.
Bottom: A \textit{double} twist, a full turn of the (red) segment, is topologically equivalent, modulo strand rearrangements, to no twist at all.
As a consequence, as shown in the text,
twists generate the Lie group U(1).}

\section{U(1) and the first Reidemeister move} 
\label{sec:u1}

\noindent
The first type of fermion core deformations of interest in quantum theory is the first Reidemeister move, the (full) \textit{twist}, which is illustrated in \figureref{ill-twist-u1}.
In three dimensions, a (full) twist can be described as a rotation of a strand 
\textit{segment} by $\pi$ around a given axis.
If desired, one can imagine that the dotted circle in \figureref{ill-twist-u1} is a fist grabbing the segment from the left, or that it 
is the border of a transparent plastic disc on which the strand segment is glued. 
Such a localized strand segment rotation by $\pi$ behaves like the generator of U(1), as shown in the following. The arguments about twists are 
a warm-up for the cases of pokes and slides.

In three dimensions, a \textit{crossing} is an \textit{observer-dependent} feature.
For a small set of observers, located in the plane defined by the straight strand and the rotation axis, the rotation of a strand segment by $\pi$ around the axis does \textit{not} generate a crossing, and thus does \textit{not} generate a twist. %
Now, the basis of the strand tangle model is that only crossing switches are observable, i.e., only the disappearance 
of a crossing with one sign and the appearance of a crossing with the opposite sign are observable. 
This implies that the mentioned small set of observers that do not observe crossings make no observation at all. 
In addition, due to the continuous fluctuations of strand shapes, the small set of observers is negligibly small; 
it thus can safely be ignored.
The overwhelming majority of observers will observe a crossing when a twist move is performed.
And as a consequence of the fundamental principle of the strand tangle model, \textit{twists, or first Reidemeister moves, are observable.}
And indeed, every twist on a fermion core yields a change in its phase.


The twist illustrated in \figureref{ill-twist-u1} can be generalized to \textit{arbitrary} angles: one can imagine that the strand segment enclosed 
by the dotted circle  -- or, if desired, the transparent plastic disc containing it -- is rotated 
by an \textit{arbitrary} angle around the rotation axis. (Sometimes an additional shift of the segment perpendicularly to the tethers is helpful for visualization.)
Such a local deformation is best called a \textit{generalized twist}.
(Generalized twists do not produce crossings for all observers.)

Generalized twists on a given strand segment around a given axis obey the \textit{group axioms:} 
they can be concatenated (multiplied), the concatenation is associative, there is a neutral 
element (no rotation at all), and every generalized twist has an inverse (the inverse rotation 
of the segment). 
Thus, generalized twists form a group. 
Generalized twists are parametrized by a single real angle: they form a manifold.
Concatenations behave nicely on this manifold. 
Therefore, generalized twists form a \textit{one-dimensional Lie group.} 
The concatenation of generalized twists is also \textit{commutative}.
Above all, as shown in Figure 5, the concatenation of two (full) twists is topologically equivalent, modulo 
strand rearrangements, to no twist at all. Full twists thus behave like rotations by $\pi$ on 
a circle: two such rotations are equivalent, modulo $2 \pi$, to no rotation.
The group defined by generalized twists is thus \textit{compact}.
The only one-dimensional, commutative, and compact Lie group is the circle group U(1),
also denoted $S^1$.
U(1) is also the set of unit complex numbers.

Recapitulating, the description of a fermion with a tangle implies that gauge interactions are due to strand deformations and change the total phase of the fermion. 
Twists are a class of strand deformations and form a group. 
The filiform structure of strands implies that the representations of their deformation group are unitary.
Modelling gauge interactions with deformations of strands thus explains the ‘U’ of U(1). 
The single strand involved in twists explains the ‘(1)’ of U(1).

In short, generalized twists for a given strand segment around a given axis form the Lie group U(1). 
The first Reidemeister move, the (full) twist, acts as the \textit{generator} of the group U(1) of generalized twists. 
In the strand description of wave functions, particles and interactions, the twist plays an important role.
A twist changes the phase of a fermion and thus models an interaction.
A separate publication has shown that \figureref{ill-gauge} and \figureref{ill-twist-u1} can be used to define a complete model for the electromagnetic 
interaction, for the electric charge, for the photon,
for the electromagnetic coupling constant, and for perturbative quantum electrodynamics \cite{csqed}. 
Twists in strands predict that no measurable deviation from quantum electrodynamics will ever be observed. 


\incepsfig{ill-poke-su2}{1}{%
Top: the three types of (full) pokes -- second Reidemeister moves -- are illustrated.
Pokes are most practically described by rotating by $\pi$ around the rotation axis a dotted
region -- or a transparent plastic disc -- containing \textit{two} strand segments.
The middle image illustrates the textbook form of the poke. 
The pokes for the other two rotation axes are illustrated on the outer sides.
Bottom: \textit{fourfold} pokes -- i.e., two turns of the plastic 
disc (or belt buckle) --
are topologically equivalent, modulo strand rearrangements, to no poke at all. 
This is the \textit{belt trick.}
As a result, as shown in the text,
pokes generate the Lie group SU(2).}

\section{SU(2) and the second Reidemeister move} 
\label{sec:su2}

\noindent
The next type of fermion core deformation of interest in quantum theory is the second Reidemeister move, the (full) \textit{poke}, which is a local rotation of a 
region enclosing \textit{two} strand segments by $\pi$,  as illustrated in \figureref{ill-poke-su2}. 
Three linear independent local rotations are possible, along the three axes that are perpendicular to each other.
The three pokes are called 
$\tau_{x}$, $\tau_{y}$ and $\tau_{z}$.
One can visualize each poke with the rotation of a transparent disc containing the two strand segments.
The transparent disc behaves like a belt buckle. 
These three types of pokes behave like the generators of SU(2), as shown in the following. The arguments are well-known from the belt trick -- also called the scissor trick, plate trick, Dirac trick, or quaternion demonstrator. 
The arguments are given in a way that is also helpful for the next section, which will explore slides.

%
The basis of the strand tangle model is that only crossing switches are observable, i.e., only  
exchanges of a crossing with a crossing of opposite sign are observable. 
It is worth recalling that in three dimensions, crossings are (slightly) {observer-dependent} features.
A very small set of observers of the deformations illustrated in  \figureref{ill-poke-su2}, located in a plane whose location depends on the specific rotation by $\pi$, observe \textit{no} crossing. %
Again, due to the continuous fluctuations of strand shapes, the very small set of observers can be neglected.
The overwhelming majority of observers will observe crossings when a poke is performed.
As a consequence of the fundamental principle of the strand tangle model, \textit{pokes, or second Reidemeister moves, are observable.} 
Specifically, every poke in a particle core yields a change in the phase of the particle.

The mathematical properties of pokes can be explored with an analogy from the human body. 
One's hand can be taken as the circled region, and the arm represents three or more tethers. 
If desired, one can take two hands holding each other, to represent even more tethers.
Exploring the behaviour of the tethered circled region -- or transparent plastic disc, or belt buckle, or hand -- defined in \figureref{ill-poke-su2}, one finds several results. 
First, concatenating two pokes by the angle $\pi$ around two perpendicular axes yields the third -- or its negative. 
Second, the concatenation of different pokes around two perpendicular axes anti-commutes.
%

Finally, one finds that for all three pokes, \textit{the fourth power is the identity.}
This result, illustrated in \figureref{ill-poke-su2}, is the {belt trick} used by Dirac.
The belt trick shows that a tethered object -- such as a hand, a belt buckle or the plastic disc in the figure enclosing two strand segments -- 
returns to its previous situation every \textit{two turns.}
An animation of the belt trick by Antonio Martos 
and animations for various numbers of tethers by Jason Hise are found at \url{https://www.motionmountain.net/videos.html#strands}.
%
The fourth power of each poke, a rotation by $4\pi$, is the identity, and the square of each poke, a rotation by $2\pi$, is~$-1$.
This applies to any belt buckle or structure with more than two tethers (or with at least one belt). 
%
All properties can be summarized in the following concatenation 
(multiplication) table for the generators
\begin{equation}
 \let\l\tau 
 \def\mmm{\phantom{-}}
\begin{array}{c|ccc} 
  \cdot & \mmm\l_{x}    &    \mmm\l_y &    \mmm\l_z  \\  
\hline 
\l_{x}    &   -1    &  -\l_z  & \mmm\l_y   \\ 
\l_y    & \mmm\l_z  &  -1     &  -\l_{x}   \\ 
\l_z    & -\l_y  & \mmm\l_{x}  &   -1     \\ 
\end{array}
\label{weaktab}
\end{equation}
In this table, the entry $-1$ means that the belt buckle (or the dotted circle, 
or the hand) has rotated by $2\pi$ and that the strand tethers (or the arm) are twisted. 
One notes that the operators $\tau_n$ behave like $i$ times the Pauli matrices:
\begin{equation}
 \let\l\tau 
  \l_{x}=i\sigma_{x} = i
    \begin{pmatrix}
      0&\phantom{-}1\\
      1&\phantom{-}0
    \end{pmatrix} \ , \ 
  \l_{y}=i\sigma_{y} = i
    \begin{pmatrix}
      0& -i \\
      i&\phantom{-}0
    \end{pmatrix} \ , \ 
  \l_{z}=i\sigma_{z} = i
    \begin{pmatrix}
      1&\phantom{-}0\\
      0&-1
    \end{pmatrix} 
\end{equation}
Using the relation that $i$ corresponds to a rotation by $\pi$, one can deduce the matrix entries in the poke matrices directly from the behaviour of the two strand segments illustrated in \figureref{ill-poke-su2}: each entry tells whether an encircled segment is rotated or not, whether it switched position with the other one or not, and how the segments switched.

In total, all the properties deduced from \figureref{ill-poke-su2} prove that the three pokes generate the Lie algebra of SU(2).
If one prefers, one can deduce the Lie algebra of SU(2) also by introducing the \textit{commutator} between pokes. For example, this can be done by using the matrix representations.

The step from the Lie algebra to the Lie group arises by generalizing the three pokes to arbitrary angles: one can imagine that the dotted circles
containing the two segments 
in \figureref{ill-poke-su2} are rotated by an \textit{arbitrary} angle around the rotation axes.
Such deformations are best called \textit{generalized pokes}.
(Also generalized pokes produce crossings only for some observers.)

Generalized pokes obey the group axioms: they can be concatenated (multiplied), the concatenation is associative, 
there is a neutral element (no rotation at all), and 
each generalized poke has an inverse (the inverse rotation of the segments). 
Generalized pokes thus form a \textit{group.}
Generalized pokes are parametrized by three real angles,
they form a manifold, and their multiplication behaves nicely on this manifold.
Therefore, generalized pokes form a \textit{three-dimensional Lie group}.
Being described by angles, the Lie group defined by generalized pokes is also \textit{compact}. 
Together with the Lie algebra of the generators, this implies: \textit{generalized pokes form the Lie group SU(2).}
The Lie group SU(2) is also the set of unit quaternions.


The description with tether deformations, the unitarity of the group elements, and the Hermitian property of the Pauli matrices all imply each other. 
Modelling gauge interactions with strand deformations thus explains the `U' of SU(2).
Without the tethers, unitarity would not arise. 
Without tethers, the group arising from the rotations of the dotted circle would be orthogonal, not unitary, and would be SO(3).
The impenetrability of strands implies vanishing traces of the representing matrices, implies the determinant $+1$, and explains the `S' of SU(2).
Finally, the two strands involved in pokes explain the `(2)' of SU(2).

It should be remarked that generalized pokes can also be described by the deformations of a \textit{single} strand 
segment -- i.e., with a plastic disc containing just one strand segment --
\textit{provided} that the deformations with respect to the other strand are 
defined properly. 
One then needs to combine a \textit{rotation} of the disc with a \textit{linked translation} with respect to the other strand: 
this is a screw-like motion of the strand segment. (In a screw, rotation and translation are linked.)
The results for \textit{single} strand deformations are the same as those just derived for strand \textit{pair} deformations. 
The description using deformations of strand {pairs} was chosen above because the analogy with the belt trick is more intuitive, 
and because it is helpful in the next section.

In short, the second Reidemeister move, the poke, yields three deformations that generate the Lie algebra SU(2). 
This is as expected and known from the belt trick.
Generalized pokes form the full Lie group SU(2).
In the strand description of wave functions, particles and interactions, pokes play an important role.
A poke changes the phase of a fermion and thus models an interaction.
Pokes and their Lie group SU(2) can be used to define a model for the weak interaction and for the weak bosons (before symmetry breaking).
Also parity violation, the mixing with quantum electrodynamics, symmetry breaking, and all other effects of the weak interaction arise 
naturally in the strand tangle model, including the weak coupling constant,  the mixing angles and the quantum field theory of the weak interaction \cite{cspepan,csorigin}. 
Pokes in strands predict that no deviation from the usual description of the weak interaction -- called \textit{quantum asthenodynamics} by Weisskopf -- will ever be observed. 

\newpage

\incepsfig{ill-slide-su3}{1}{%
Three types of the (full) third Reidemeister moves -- three types of slides -- are illustrated.
In this triplet, each move deforms a pair of crossing strands and leaves one (black) strand undeformed.
Each move can be seen as a rotation of the region inside the dotted circle plus a related shift of that 
circle in the direction of the central starting triangle, in the way of the motion of a screw.
The three moves generate an SU(2) (sub)algebra that corresponds to the belt trick for the two crossing strands.
Two further triplets of slides are illustrated in \protect\figureref{ill-gluon}.}
\incepsfig{ill-gluon}{1}{The ten important deformations 
deduced from the third Reidemeister move, the \textit{slide}, are illustrated.
(The graphs are squashed vertically to save space.)
Each deformation, apart from $\lambda_{8}$, rotates a dotted circle by $\pi$.
Each of the upper three rows defines an SU(2) subgroup.
Using the definition of $\lambda_{8}$ given in the text,
the eight slides $i\lambda_{1},\ldots,i\lambda_{8}$, thus without 
$i\lambda_{9}$ and $i\lambda_{10}$, turn out to generate the Lie group SU(3).
The corresponding Gell-Mann matrices are given in \protect\tableref{tabmat}.}[%
\psfrag{t1}{\small $t$}%
\psfrag{t2}{\small $t+\Delta t$}%
]

\section{SU(3) and the third Reidemeister move} 
\label{sec:su3}

\noindent
The last type of fermion core deformations of interest in quantum theory is the third Reidemeister move, the (full) \textit{slide}, which occurs in configurations in which three strands are on top of each other,
as illustrated in \figureref{ill-slide-su3}. 
In particular, a \textit{slide} can be described as a deformation of a (dotted circle) 
region of \textit{two} crossing strand segments against a third strand: the deformation consists of a rotation of the region by $\pi$ and a subsequent linked shift towards the third strand. Rotations and shifts are linked as in the motion of a screw. 
Three such local rotations of strand pairs by $\pi$ and linked shifts are illustrated in the figure.

The basis of the strand tangle model is that only crossing switches are observable, i.e., only the 
disappearance of a crossing and the subsequent appearance of a crossing with the opposite sign are observable. 
As a consequence of the fundamental principle of the strand tangle model, also third Reidemeister moves, or slides, are observable. 
Every slide yields a change in the phase of the fermion tangle.

The three deformations of \figureref{ill-slide-su3} are linearly independent, as they take place along mutually perpendicular axes.
Imagining a transparent plastic disc glued to the crossing strands inside the dotted circle helps to visualize the deformations.
Like in the situation described in the previous section, on pokes, the circled region behaves like the belt buckle in the belt trick. 
The three deformations of \figureref{ill-slide-su3} thus act as generators of an SU(2) subalgebra.

In a situation with three strands, \textit{for each strand pair} there are three different rotation-shifts.
This yields a total of \textit{nine} deformations. 
The nine deformations are illustrated in the upper three rows of \figureref{ill-gluon}.
All deformations in the figure are called \textit{full slides} in the following.
As shown next, the nine types of deformations just defined can be combined into \textit{eight} 
linearly independent ones that behave like the generators of SU(3).
This is most simply done by defining an additional deformation that is illustrated at the bottom of \figureref{ill-gluon}, bringing the total to ten slides.
To show that SU(3) arises from \figureref{ill-gluon}, the Gell-Mann matrix representation 
of \tableref{tabmat} and the multiplication \tableref{su3mutablambda} are deduced in the following.
They imply the Lie algebra of SU(3).
Afterwards, the Lie group is deduced.

\figureref{ill-gluon} implies that the slides $\lambda_3$, $\lambda_9$, and $\lambda_{10}$ are \textit{linearly dependent.}
Indeed, these three deformations act in related ways, in the same plane, on the three strands. 
The figure shows that only two of three slides $\lambda_3$, $\lambda_9$ and $\lambda_{10}$ are linearly independent:
two types of slides are sufficient to move all three strands.
To have an orthonormal basis, it is customary to use the deformation $\lambda_3$ and the 
additionally defined deformation $\lambda_{8}=(\lambda_{10}-\lambda_{9})/\smallsqrt{3}$. 
The factor $\smallsqrt{3}=2\, \sin (2\pi/3)$ is due to the angle $2\pi/3$ that describes the 
threefold axis at the centre of the three-strand configuration.
As a consequence of this threefold symmetry, the square root of three appears in many places in SU(3).

Using the definition of $i\lambda_{8}$, the eight slides $i\lambda_{1},\ldots,i\lambda_{8}$ illustrated in \figureref{ill-gluon} are all linearly independent of each other. 
In particular, the figure illustrates that $i\lambda_{1},\ldots,i\lambda_{7}$ either act on different dotted circles or at least act along linearly independent directions.
The generator $i\lambda_{8}$ is special.
Due to its definition, $i\lambda_{8}$ is linearly independent of the first seven slides, and it is the only full slide that deforms all \textit{three} strands.
At the same time, $i\lambda_{8}$
is the slide that resembles most the original definition of the third Reidemeister move,
which was illustrated on top of \figureref{ill-slide-su3}. 
In contrast to the upper nine deformations in \figureref{ill-gluon} that seem 
so different from the third Reidemeister move, $i\lambda_{8}$ confirms that the slides indeed are generalizations of the third Reidemeister moves.

\begin{table}[t]
\small
\caption{The matrix representations for the Hermitian operators $\lambda_1...\lambda_{10}$ are listed. 
As shown in the text, the representations follow from the moves illustrated in \figureref{ill-gluon}.
The eight SU(3) generators are given by $i\lambda_{1},\ldots,i\lambda_{8}$.
For these eight operators, the definition of $\lambda_8$ yields the general trace relations $\operatorname{tr}\lambda_n = 0$ and $\operatorname{tr}(\lambda_n \lambda_m) = 2\delta_{nm}$.
This particular matrix representation is called the Gell-Mann representation.}%
\label{tabmat}  
\def\mmm{\phantom{-}}
\begin{eqnarray}
\lambda_1 = \begin{pmatrix} \mmm0 & \mmm1 & \mmm0 \\ \mmm1 & \mmm0 & \mmm0 \\ \mmm0 & \mmm0 & \mmm0 \end{pmatrix}, \ \ \ \  
\lambda_2 = \begin{pmatrix} \mmm0 & - i & \mmm0 \\  \mmm i & \mmm 0 & \mmm 0 \\ \mmm 0 & \mmm 0 & \mmm 0 \end{pmatrix}, \ \ \ \ 
\lambda_3 = \begin{pmatrix} \mmm1 & \mmm0 & \mmm0 \\ \mmm0 & -1 & \mmm0 \\ \mmm0 & \mmm0 & \mmm0 \end{pmatrix},
\notag
\\
\lambda_4 = \begin{pmatrix} \mmm0 & \mmm0 & \mmm1 \\ \mmm0 & \mmm0 & \mmm0 \\ \mmm1 & \mmm0 & \mmm0 \end{pmatrix}, \ \ \ \ 
\lambda_5 = \begin{pmatrix} \mmm0 & \mmm0 & - i \\ \mmm0 & \mmm0 & \mmm0 \\  \mmm i & \mmm0 & \mmm0 \end{pmatrix}, \ \ \ \ 
\lambda_9 = \begin{pmatrix} -1 & \mmm0 & \mmm0 \\ \mmm0 & \mmm0 & \mmm0 \\ \mmm0 & \mmm0 & \mmm1 \end{pmatrix},
\notag
\\
\lambda_6 = \begin{pmatrix} \mmm0 & \mmm0 & \mmm0 \\ \mmm0 & \mmm0 & \mmm1 \\ \mmm0 & \mmm1 & \mmm0 \end{pmatrix}, \ \ \ \ 
\lambda_7 = \begin{pmatrix} \mmm0 & \mmm0 & \mmm0 \\ \mmm0 & \mmm0 & - i \\ \mmm0 &  \mmm i & \mmm0 \end{pmatrix}, \ \ \ 
\lambda_{10} = \begin{pmatrix} \mmm0 & \mmm0 & \mmm0 \\ \mmm0 & \mmm1 & \mmm0 \\ \mmm0 & \mmm0 & -1 \end{pmatrix},
\notag
\\
\hbox{and} \ \ 
\lambda_8 = \frac{1}{\sqrt{3}} \begin{pmatrix} \mmm1 & \mmm0 & \mmm0 \\ \mmm0 & \mmm1 & \mmm0 \\ \mmm0 & \mmm0 & -2 \end{pmatrix}.
\notag
\end{eqnarray}
\end{table}

To deduce the matrix representation of slides given in \tableref{tabmat} and the multiplication \tableref{su3mutablambda} from \figureref{ill-gluon},
it is convenient to start with the 
triplet $i\lambda_{1}$, $i\lambda_{2}$ and $i\lambda_{3}$.
As mentioned, the corresponding slides, which were also illustrated in \figureref{ill-slide-su3}, generate an SU(2) subgroup because they deform a pair of crossing strand segments in a way that reproduces the belt trick.
The dotted circle plays the role of the belt buckle in the belt trick.
\figureref{ill-gluon} illustrates that the same happens for the other two triplets: 
each triplet is based on one crossing strand pair.
In the matrix representation of the slides given in \tableref{tabmat}, the undeformed 
strand corresponds to the column and the row containing only zeros. 
As just explained, \figureref{ill-gluon} implies that the squares $(i\lambda_{1})^2$, $(i\lambda_{2})^2$ and $(i\lambda_{3})^2$
have the diagonal values $(-1,-1,0)$ and zero everywhere else: the belt trick acts only on the first strand pair.
Using the commutation properties of the first triplet, again due to the belt trick,
one thus finds that the first two rows and columns of the first triplet reproduce the Pauli matrices. 
The matrix representation of the first triplet is thus fixed.

It is straightforward to confirm from \figureref{ill-gluon} that the concatenation of generalized slides from the same SU(2) triplet, but around different axes, is \textit{anti-commutative}, as expected from SU(2) generators.
The multiplication behaviour of the first slide triplet is listed in the multiplication 
\tableref{su3mutablambda} in the nine fields on the top left.
\figureref{ill-slide-su3} also illustrates a difference to the usual SU(2) multiplication table.  
The square of the generators $i\lambda_{1}$, $i\lambda_{2}$ and $i\lambda_{3}$ cannot be $-1$, 
because one strand remains undeformed, 
and because the crossing of the other two strands shifts \textit{against} 
the undeformed strand when the deformations are performed twice.  
Therefore, the square of each generator is effectively equal to $-1$ 
\textit{only} for the two strands that are deformed, but not for the whole structure. 
The diagonal values in the multiplication table follow once the matrix representation is complete, and the definition of $\lambda_8$ is used.

The two additional slide triplets, also forming SU(2) subgroups, are illustrated in \figureref{ill-gluon}.
Their matrix representations, given in \tableref{tabmat}, follow when the corresponding strand pairs are taken into account.
The matrix representations are Pauli matrices for those strand pairs that are being deformed.
The undeformed strand in each triplet yields a vanishing column and row in the matrices.
In the matrix representation, the squares of the last two triplets thus have the diagonals $(1,0,1)$ and $(0,1,1)$ and have zero everywhere else, as expected.
In the multiplication \tableref{su3mutablambda}, the triplets are separated by vertical lines and by thicker horizontal lines.
Within each triplet, the squares for each slide are all equal, as expected. 

The matrix representations for the triplet $i\lambda_{1},i\lambda_{2},i\lambda_{3}$,
for $i\lambda_{4},i\lambda_{5},i\lambda_{9}$,
and for $i\lambda_{6},i\lambda_{7},i\lambda_{10}$ are thus fixed by \figureref{ill-gluon}.
This correspondence also fixes the matrix representation for $\lambda_{8}$, from its definition, 
and for its square.
In particular, all slide matrices have trace 0.
As expected, the matrices of the first eight slides are linearly independent.
The matrix representation for $\lambda_3$ and $\lambda_8$ shows that the trace of their squares 
is 2 in both cases, that they commute, and above all, that they are orthogonal to each other.
In short, the trace of slide products is ${\rm tr}(\lambda_{n}\lambda_{m})=2 \delta_{nm}$.

{\footnotesize
\begin{table}[htp]
\small
\centering
\caption{As shown in the text, \figureref{ill-gluon} implies the following 
multiplication table, using the concatenation of deformations as multiplication.
Barring the rows and columns for $\lambda_{9}$ and $\lambda_{10}$,
and multiplying each slide $\lambda_{n}$ by $i$, yields the multiplication
table of the generators of SU(3).  
The \emph{linearly dependent} slides 
$\lambda_{9}=-\lambda_{3}/2 - \lambda_{8} \smallsqrt{3}/2$ and
$\lambda_{10}=-\lambda_{3}/2 + \lambda_{8} \smallsqrt{3}/2$
do \emph{not} yield generators.
These two slides are used to construct
$\lambda_{8}$ using $\lambda_{8}=(\lambda_{10}-\lambda_{9})/\smallsqrt{3}$.%
\label{su3mutablambda}
The three SU(2) subgroups are generated by the triplet 
$\lambda_{1}$, $\lambda_{2}$, and $\lambda_{3}$, the triplet
$\lambda_{4}$, $\lambda_{5}$ and $\lambda_{9}$, and the triplet
$\lambda_{6}$, $\lambda_{7}$, and $\lambda_{10}$.
Despite the first impression, $\lambda_{4}^2=\lambda_{5}^2=\lambda_{9}^2$
and $\lambda_{6}^2=\lambda_{7}^2=\lambda_{10}^2$.}
\hspace*{-28mm}\vbox{\setlength\textwidth{162mm}\begin{equation*}
\setlength\textwidth{162mm}%
\small%
\scriptstyle%
\small%
\def\s{\smallsqrt{3}}\let\l\lambda%
\begin{array}{@{\hspace{0em}}c|c@{\hspace{0em}}c@{\hspace{0em}}c|c@{\hspace{0em}}%
c@{\hspace{0em}}c|c@{\hspace{0em}}c@{\hspace{0em}}c|c@{\hspace{0em}}}%
&  \l_1  &  \l_2  &  \l_3  &  \l_4   &  \l_5     &  \l_9&  \l_6  &  \l_7  &  \l_{10} &  \l_8  \\ %
\toprule 
\l_1\cstabhlineup %
&   2/3  &  i\l_3 & -i\l_2 & \l_6/2 &  -i\l_6/2 &-\l_1/2& \l_4/2&-i\l_4/2& \l_1/2&\l_1/\s\\ %
\small %
\ \cstabhlinedown %
&+\l_8/\s&        &        &+i\l_7/2 &   +\l_7/2 &+i\l_2/2&+i\l_5/2 &+\l_5/2&+i\l_2/2&\\ %
\hline 
\l_2\cstabhlineup %
& -i\l_3 &    2/3 &  i\l_1 & i\l_6/2 &   \l_6/2 &-i\l_1/2&-i\l_4/2&-\l_4/2&-i\l_1/2&\l_2/\s\\ %
\ \cstabhlinedown %
&        &+\l_8/\s&        & -\l_7/2 &  +i\l_7/2  &-\l_2/2& +\l_5/2&-i\l_5/2 &+\l_2/2&\\ %
\hline 
\l_3\cstabhlineup %
&  i\l_2 & -i\l_1 &    2/3 & \l_4/2 &  -i\l_4/2&-1/3-\l_3/3&-\l_6/2&i\l_6/2 &-1/3+\l_3/3&\l_3/\s\\ %
\ \cstabhlinedown %
&        &        &+\l_8/\s& +i\l_5/2 &   +\l_5/2 &+\l_9/3&-i\l_7/2 &-\l_7/2&+\l_{10}/3&  \\ %
\toprule 
\l_4\cstabhlineup %
&\l_6/2&-i\l_6/2&\l_4/2   &2/3+\l_3/2& -i\l_9 & i\l_5 &\l_1/2 &i\l_1/2 &-\l_4/2 &-\l_4/2\s \\ %
\ \cstabhlinedown %
&-i\l_7/2&-\l_7/2&-i\l_5/2&-\l_8/2\s &         &      &+i\l_2/2 &-\l_2/2 &-i\l_5/2&-i\s\l_5/2\\ %
\hline 
\l_5\cstabhlineup %
& i\l_6/2&\l_6/2&i\l_4/2 &   i\l_9  &2/3+\l_3/2&-i\l_4&-i\l_1/2&\l_1/2 &i\l_4/2&i\s\l_4/2 \\ %
\ \cstabhlinedown %
& +\l_7/2&-i\l_7/2&+\l_5/2 &        &-\l_8/2\s &      &+\l_2/2 &+i\l_2/2 &-\l_5/2&-\l_5/2\s\\ %
\hline 
\l_9\cstabhlineup %
&-\l_1/2 &i\l_1/2&-1/3-\l_3/3&-i\l_5&i\l_4&2/3+2\l_3/3&\l_6/2  &i\l_6/2&-1/3-\l_9/3& -1\\ %
\ \cstabhlinedown %
&-i\l_2/2&-\l_2/2&+\l_9/3    &      &     &+\l_9/3&-i\l_7/2&+\l_7/2&+\l_{10}/3& +\l_{10}\\ %
\toprule 
\l_6\cstabhlineup %
&+\l_4/2&i\l_4/2 &-\l_6/2 &   \l_1/2  &i\l_1/2&\l_6/2&2/3-\l_3/2 &i\l_{10}&-i\l_7&- \l_6/2\s\\ %
\ \cstabhlinedown %
&-i\l_5/2&+\l_5/2 &+i\l_7/2 & -i\l_2/2 &+\l_2/2&+i\l_7/2&-\l_8/2\s&        &       &-i\s\l_7/2 \\ %
\hline 
\l_7\cstabhlineup %
&i\l_4/2 &-\l_4/2 &-i\l_6/2&  -i\l_1/2&  \l_1/2&-i\l_6/2&-i\l_{10}& 2/3-\l_3/2&i\l_6&i\s\l_6/2 \\ %
\ \cstabhlinedown %
&+\l_5/2 &+i\l_5/2 &-\l_7/2 &  -\l_2/2 &-i\l_2/2&+\l_7/2&         &-\l_8/2\s&     &-\l_7/2\s\\ %
\hline
\l_{10}\cstabhlineup %
&-\l_1/2&-i\l_1/2&-1/3+\l_3/3&-\l_4/2&-i\l_4/2&-1/3-\l_9/3&i\l_7&-i\l_6 &2/3-\l_3/3&1\\ %
\ \cstabhlinedown %
&+i\l_2/2&-\l_2/2&-\l_{10}/3&+i\l_5/2&-\l_5/2&+\l_{10}/3 &      &       &+\l_9/3   &+\l_9\\ %
\toprule 
\l_8\cstabhlineup %
&\l_1/\s &\l_2/\s& \l_3/\s&-\l_4/2\s&-i\s\l_4/2&-1      &-\l_6/2\s&-i\s\l_6/2& 1    &2/3 \\ %
\ \cstabhlinedown %
&        &       &        &+i\s\l_5/2&- \l_5/2\s&+\l_{10}&+i\s\l_7/2&-\l_7/2\s& +\l_9&-\l_8/\s\\ %
\end{array}%
\nonumber
\end{equation*}}\hss%
\end{table}}%

In short, the preceding arguments prove that \figureref{ill-gluon} fixes the specific matrix representation of the deformations $i\lambda_{1},\ldots,i\lambda_{8}$ that is given in \tableref{tabmat}.
This matrix representation is called the \textit{Gell-Mann representation}.
The resulting multiplication table is given in \tableref{su3mutablambda}.
The matrix representation defines the SU(3) Lie algebra.
Together, the matrix representation and the multiplication table also yield the matrix commutators.
The commutators confirm that the compact, non-commutative, eight-dimensional Lie algebra defined by the eight generators $i\lambda_{1},\ldots,i\lambda_{8}$, with its three SU(2) subalgebras, is the standard SU(3) Lie algebra.
The SU(3) structure constants can be deduced from the matrix representation.

To get the SU(3) Lie group from the SU(3) Lie algebra, like for the simpler gauge groups, the 
deformations of \figureref{ill-gluon}
must be generalized to arbitrary angles: 
one can imagine that the crossings inside the dotted circles 
are deformed by an arbitrary angle around the shift-rotation axis.
Such deformations are called \textit{generalized slides} in the following.

Generalized slides obey the group axioms and form a \textit{group:} they can be concatenated (multiplied), 
the concatenation is associative, there is a neutral element (no rotation at all), 
and each generalized slide has an inverse (the inverse rotation of the segment pair).
More precisely, generalized slides form a \textit{Lie group:} they form a compact manifold,
because their parameters are (real) angles
and their concatenations behave nicely on this manifold.
The full set of generalized slides is parametrized by \textit{eight} angles or real numbers. 
The Lie group of generalized slides is thus \textit{eight-dimensional}.
Together with the SU(3) Lie algebra, all these properties imply that generalized slides defined with \figureref{ill-gluon} form the Lie group SU(3).

Also for slides, the description with deformations, the unitarity property of the group elements, and the Hermitian property of the Gell-Mann matrices all imply each other. 
Modelling gauge interactions with deformations thus explains the `U' of SU(3).
Without the tethers, unitarity would not arise.
The impenetrability of strands implies vanishing traces of the representing matrices, implies the determinant $+1$, and thus explains the `S' of SU(3).
The three strands involved in the slides explain the `(3)' of SU(3).

\section{Checking the derivation of SU(3)} 
\label{sec:su3checks}

\noindent
There are several ways to check that the deduction of SU(3) from \figureref{ill-gluon} is correct.
First of all, \figureref{ill-gluon} contains three SU(2) subalgebras, due to the three belt buckles that are contained in it.
The three SU(2) subalgebras are rotated by the angle $\pm2\pi/3$ with respect to each other, around the axis defined by the direction of observation. 
Strands thus illustrate the threefold $\rm C_3$ \textit{symmetry} of SU(3), as expected. 
(Due to the squashing of the graphs in \figureref{ill-gluon}, the symmetry is not fully obvious. It is more obvious in \figureref{ill-slide-su3}.)
The four slides on the rightmost column of \figureref{ill-gluon}
are all part of the \textit{centre} of SU(3).
In particular, the linear dependent slides $\lambda_3$, $\lambda_9$, and $\lambda_{10}$ illustrate the $\rm C_3$ symmetry of the centre of SU(3) -- again as expected.

Secondly, in each triplet, the squared slides leave one strand undeformed and shift the crossing of the other two strands towards the undeformed strand.
Indeed, in the multiplication table for the first triplet, the square of each slide involves $\lambda_{8}$.
Without the shift, $\lambda_{8}$ would not arise in the table.
Likewise, the squares in the other two triplets involve $\lambda_{3}$ and $\lambda_{8}$.
This is as expected from the threefold symmetry of SU(3).
The multiplication behaviour is the same as for the first triplet, with the diagonal products transformed by a rotation by $\pm2\pi/3$. 
Again, the result is as expected.


Thirdly, the product value of a slide with itself can be checked using \figureref{ill-gluon}.
This requires the definition of scalar multiplication and addition.
This step was not necessary in the case of pure SU(2) and is the reason that the strand realization of SU(3) was overlooked for a long time.

In the strand tangle model, the \textit{addition} of strand-pair 
deformations is realized when 
the two deformations are applied at \textit{the same time}.
The scalar multiplication of a strand deformation is realized by \textit{multiplying} 
the corresponding \textit{rotation angle} of the circled region (the belt buckle).
(The mentioned equivalence between $2\pi$ and $-1$, and between $
\pi$ and $i$ are used.)
%
As an example, the definition $\lambda_{8}=(\lambda_{10}-\lambda_{9})/\smallsqrt{3}$ 
yields the deformation illustrated at the very bottom of \figureref{ill-gluon}: 
above all, it switches the orientation of the central triangle. 

Using the definitions of addition and scalar multiplication,
the products of each slide with itself can be checked.
For the products on the diagonal of the first triplet -- given in \figureref{ill-slide-su3} 
and in \figureref{ill-gluon} -- the matrix multiplication yields 
$\lambda_1^{2}=\lambda_2^{2}=\lambda_3^{2}= 2/3 + \lambda_8 / \smallsqrt{3} 
= 2/3 - \lambda_{9}/3 + \lambda_{10}/3$.
The numbers are \textit{not} deduced in a straightforward way directly from the deformations
in \figureref{ill-gluon}; but the deformations do show that the squares of 
the slides in the first triplet are independent of  $\lambda_3$.
Therefore the squares of the first triplet are a linear combination of the identity and $\lambda_8$.
(This linear combination narrows the central triangle between the three 
strands along the west-east direction and leaves the opposite strand untouched.)
The numbers in the definition of $i\lambda_{8}$ -- equivalently, the numbers in its diagonal matrix representation -- explain the three entries on the diagonal 
of the multiplication table for the first slide triplet. 
The values of the diagonal of the multiplication table for the other two triplets follow
after rotation by $\pm2\pi/3$ around the direction of observation.
The square of $\lambda_8$ follows.
In other words, the squares of all slides are fixed by \figureref{ill-gluon}.

Finally, \figureref{ill-gluon} allows several additional checks of the slide multiplication table.
\figureref{ill-gluon} implies that the four slides $i\lambda_3$, $i\lambda_9$, $i\lambda_{10}$ and $i\lambda_8$ in the last column all commute among each other.
This is reproduced in the multiplication table.

Compared to SU(2), which is anti-commutative, SU(3) is more strongly non-commutative.
This is best seen in the products between the first slides from different triplets.
An example is the difference between the products $\lambda_1\lambda_4$ and $\lambda_4\lambda_1$. 
Exploring the concatenation of the corresponding slides shows that the two products are \textit{not} the negative of each other. 
This happens because each belt trick operation also \textit{shifts} the belt (the region inside the dotted circle), and the shifts destroy the anti-commutation for the cases that the two slides are from different triplets. 
Due to these shifts, the product $\lambda_1\lambda_4$ yields a linear combination of the slides of the remaining triplet; and the product differs from the product $\lambda_4\lambda_1$.
This is as expected.
Slides thus do not commute in general.
And like pokes, slides generate a Yang-Mills theory \cite{csqcd}.



As a remark, the group SU(3) can also be deduced from deformations of a \textit{single} strand segment, instead of deformations of crossing strand pairs. 
However, the images are less pedagogical.
Possibly, an even more pedagogical set of eight deformations yielding SU(3) can be found.

In contrast to strands, number fields do not explain the gauge groups: even though 
U(1) are the unit complex numbers and SU(2) are the unit quaternions, the unit octonions do not form a group and have no simple relation to SU(3).

In short, the third Reidemeister move, the slide, naturally yields eight deformations that generate the Lie group SU(3) of generalized slides.
In the strand description of wave functions, particles and interactions, slides play an important role.
A slide changes the phase of a fermion and thus models an interaction.
Slides and their Lie group SU(3) can be used to define a model for the strong nuclear interaction, for the gluons, and for the colour charge, as explained elsewhere \cite{csqcd}.
The quark model, Regge trajectories, glueballs, the lack of CP violation in the strong interaction, and the strong coupling constant arise naturally.
Strands fully reproduce quantum chromodynamics
and predict that no measurable deviation from quantum chromodynamics will ever be observed.

\section{The possible gauge groups in nature} 
\label{sec:poss}

\noindent 
Explaining the gauge groups U(1), SU(2) and SU(3) as the result of strand deformations is attractive for several reasons.
First, in the research literature, \textit{no other, ab initio explanation} of the gauge groups that agrees with all experiments -- and in particular, that does not add additional, unobserved fields -- has been published. 
So far, the strand explanation is \textit{unique} and \textit{unmodifiable}. 

Secondly, the gauge groups arise as consequences of the \textit{same idea} with which Dirac explained spin $1/2$ and fermion behaviour \cite{gardner},  with which Battey-Pratt and Racey explained the Dirac equation \cite{bpr,csqed}, and with which general relativity can be deduced \cite{csindian,csbh}. The strand explanation is \textit{simple}, \textit{consistent} and \textit{complete}.

Thirdly, the explanation of the gauge groups predicts the \textit{lack of additional gauge groups} in nature, and in particular the lack of larger, unified gauge groups. 
For example, strands imply that gauge groups like SU(5), SO(10) or E8 do not exist in nature, and neither does any other Yang-Mills theory.
Again, the explanation with strands \textit{agrees with all experiments} performed so far~\cite{pdgnew}. 
The strand tangle model is \textit{correct}.

Fourthly, in a similar way that the classification of tangle deformations 
leads to the gauge interactions, also the classification of rational tangle structures
leads to the observed \textit{elementary particles}, and to no additional ones \cite{cspepan,csorigin}.
Dark matter is predicted not to be made of unknown elementary particles.
The strand tangle model is \textit{predictive}.

Finally, strands imply that the fundamental constants -- coupling constants, mixing angles and particle mass ratios -- have \textit{unique and calculable values.}
In particular, strands imply that the statistics of their shape fluctuations allow calculating these values. 
The first rough estimates agree with data \cite{csqed,csqcd}.
The strand tangle model is \textit{testable}.


In short, the strand tangle model,
in contrast to other approaches, implies the lack of additional gauge groups in nature.
In particular, this implies the lack of a unified gauge group.
Strands also imply the lack of any other new physics.
Strands further imply unique values for the fundamental constants that are of the order of the measured values.
Due to the wide-ranging implications of the strand tangle model, a thorough check -- both mathematical and experimental -- of all its consequences should be performed.

\section{Conclusion} 
\label{sec:conc}

\noindent 
The three Reidemeister moves -- twists, pokes and slides -- have been shown to generate the Lie groups U(1), SU(2), and SU(3), once the moves are interpreted as deformations of strand tangles that model particles, wave functions and interactions.
Because Reidemeister proved that every tangle deformation is a combination of the three moves \textit{only},
the strand tangle model implies  \textit{the lack of any other gauge group in nature.}
So far, this conclusion agrees with all observations.
It appears that the explanation of gauge theory using strands and their deformations is unique, correct, simple, ab initio, consistent, complete, unmodifiable, predictive and testable.
The wide explanatory power of the strand model suggests exploring it as an approach to unification.
Experimental observation of any new physics beyond the standard model with massive Dirac neutrinos with PMNS mixing would falsify the model.
Comparing high-precision calculations of the coupling constants and the other fundamental constants to the measured values will provide definite tests of the model.

\section{Acknowledgements and declarations}  
\label{sec:ack}

\noindent
The author thanks Thomas Racey, Michel Talagrand, Jason Hise, John Baez,
Sebastian Meyer and Isabella Borgogelli for stimulating discussions and support.
This work was supported partly by a grant from the Klaus Tschira Foundation.
The author declares that he has no conflict of interest and no competing interests. 
No additional data are associated with the text.

\sloppy
\bibliography{SU3} 

\end{document}